\documentclass[12pt]{article}
\usepackage{epsfig,amssymb}

\newcommand{\be}{\begin{equation}}
\newcommand{\ee}{\end{equation}}
\newcommand{\ba}{\begin{eqnarray}}
\newcommand{\ea}{\end{eqnarray}}
\newcommand{\bq}{\begin{equation}}
\newcommand{\eq}{\end{equation}}
\newcommand{\bqa}{\begin{eqnarray}}
\newcommand{\eqa}{\end{eqnarray}}
\newcommand{\ben}{\begin{enumerate}}
\newcommand{\een}{\end{enumerate}}
\newcommand{\bc}{\begin{center}}
\newcommand{\ec}{\end{center}}
\newcommand{\bqb}{\begin{eqnarray*}}
\newcommand{\eqb}{\end{eqnarray*}}

\def\pr#1#2#3{ Phys. Rev. ${\bf{#1}}$ (#2) #3}

\def\pl#1#2#3{ Phys. Lett. ${\bf{#1}}$ (#2) #3}
\def\prep#1#2#3{ Phys. Rep. ${\bf{#1}}$ (#2) #3}

\def\np#1#2#3{ Nucl. Phys. ${\bf{#1}}$ (#2) #3}

\def\ijmp#1#2#3{ Int. J. Mod. Phys. ${\bf{#1}}$ (#2) #3}

\def\etal{{\it et.al.\/}}

\global\nulldelimiterspace = 0pt

\def\N{ {\cal N }}
\def\V{ {\cal V }}

\begin{document}
\pagenumbering{arabic}
\thispagestyle{empty}
\def\thefootnote{\fnsymbol{footnote}}
\setcounter{footnote}{1}

\begin{flushright}
PM/99-26 \\
June 1999\\
corrected version\\

 \end{flushright}
\vspace{2cm}
\begin{center}
{\Large\bf 
Logarithmic expansion of electroweak corrections \\ 
to four-fermion processes in the TeV
region
}\\
\vspace*{0.5cm}
{\large 
M. Beccaria$^{(1)}$, 
P. Ciafaloni$^{(2}$, 
D. Comelli$^{(3)}$, \\
F. Renard$^{(4)}$,
C. Verzegnassi$^{(1)}$}\\
\vspace{0.7cm}
${}^1$ 
Dipartimento di Fisica dell'Universit\`a di Lecce, I-73100, Italy,\\
and Istituto Nazionale di Fisica Nucleare, Sezione di Lecce;\\
${}^2$
Istituto Nazionale di Fisica Nucleare, Sezione di Lecce; \\
${}^3$
Istituto Nazionale di Fisica Nucleare, Sezione di Ferrara; \\
${}^4$
Physique
Math\'{e}matique et Th\'{e}orique, UMR 5825\\
Universit\'{e} Montpellier
II,  F-34095 Montpellier Cedex 5.

\vspace{0.2cm}

\vspace*{1cm}

{\bf Abstract}

\end{center}

Starting from a theoretical representation of 
the  electroweak component of 
four-fermion neutral
current processes that uses as theoretical input the experimental
measurements at the $Z$ peak, we consider the asymptotic 
high energy behaviour in
the Standard Model at one loop of 
those gauge-invariant combinations
 of self-energies, vertices and boxes that contribute all the different
observables. 
We find that the  logarithmic contribution due
to the renormalization group running of the various couplings is
numerically overwhelmed by single  and double logarithmic 
terms of
purely electroweak (Sudakov-type) origin,
whose separate relative effects grow with energy,
 reaching the 10 $\%$ size at about one TeV.
We then propose a simple
"effective" parametrization that aims at describing the various
observables in the TeV region, and discuss its validity both beyond and
below 1 TeV, in particular in the expected energy range of future
linear electron-positron (LC) and muon-muon colliders.

\def\thefootnote{\arabic{footnote}}
\setcounter{footnote}{0}
\clearpage

\section{Introduction}

\hspace{0.6cm}The construction of future 
lepton-antilepton ($l^+l^-$) colliders in
energy ranges varying from a few hundreds of GeV (LC) to a few $TeV$
(muon collider) is being thoroughly investigated at the moment
\cite{LC, muon}. One of the key points of all existing proposals is the
availability of extremely high luminosities. These would lead to an
experimental accuracy for standard four fermion processes (i.e, $l^+l^-
\to f\bar f$) comparable with that obtained at the $Z$ peak, thus
allowing high precision tests of electroweak models at one loop to be
performed in the same spirit.\par
 On the theoretical side, the possibility
that such extremely precise measurements are performed requires
imperatively the existence of suitable computational programs that are
able to provide numerical predictions of comparable accuracy, both for
the Standard Model case and, possibly, for electroweak models of
different origin. For LEP2 Physics such programs already exist at the
one loop level for the SM\cite{LEP2} and, at least in principle, their
extrapolation to higher energies should be conceivable. There are
though, in our opinion, at least two points that deserve special
attention in this respect.\par
 The first one is the importance of
understanding the role of the several terms that contribute the various
observables when the energy becomes very large. At the $Z$ peak, the
most spectacular one loop effects were provided by fermion contributions
to gauge boson
self-energies (the Higgs effect is notoriously\cite{Hscreen} screened),
with the exception of the significant  contribution due to the
celebrated $Zb\bar b$ vertex\cite{Zbb}. When the energy increases and
moves towards the $1~TeV$ region, this fermion dominance is apparently
weakened, and bosonic effects of vertex and particularly of box type
appear to rise, as first stressed in a previous paper \cite{rising}. To
confirm this rise and to understand in a simple way its physical origin
would be, in our opinion, an important achievement both within the
Standard Model framework and beyond it.\par
The second point that it might be worth to examine is that of the
reliability of a perturbative description at the one loop level when
the energy becomes very large, say beyond the $1~TeV$ orientative
value. In particular, a problem that might arise is that of having to
perform a resummation that takes into account the higher order leading
effects, like in the case of the running of $\alpha_{QED}$. In fact, it
is well-known that this problem is present in QED and QCD diagrams when
Sudakov logarithms of the type $ln q^2/\mu^2$, $ln^2 q^2/\mu^2$ (where
$\sqrt{q^2}$ is the available energy and $\mu$ is the mass scale of the
diagram) appear in vertices and boxes \cite{Sudakov}. In the Standard
Model, logarithms of similar type (usually called "of Sudakov type") are also 
known to appear\cite{Kuroda, Comelli}. In reference \cite{Comelli} the 
leading asymptotic behavior growing like  $ln^2 q^2/M_{W,Z}^2$ 
was calculated for the relevant observables. The physical  origin
of these double logs was also discussed. In this paper we also calculate 
the    
subleading terms, growing like a single   log $ln q^2/M_{W,Z}^2$,where
$M_{W,Z}$ are the physical gauge boson masses. If the relative
contribution coming from these diagrams became large, typically beyond
a "reasonable" percentage amont (that could also be quantitatively fixed
by a knowledge of the requested theoretical accuracy), the necessity of
a resummation would become stringent (and, to our knowledge, this study
has not yet been performed).\par
The aim of this paper is precisely that of discussing the two previous
points. In particular, we shall perform in Section 2 an explicit
calculation of the coefficients of the three types of logarithms that
dominate the large energy behaviour of the different one loop
gauge-invariant combinations that make up all observables of the
process. These are coming from the renormalization group (RG) running
of the gauge couplings, and from the typically electroweak Sudakov-type
effects. The first ones generate linear logarithms, the second ones
linear and quadratic logarithms. As we shall show, the numerical size
of the various Sudakov logarithms largely overwhelms that of the RG ones.
Technically speaking, the main contribution turns out to be provided by
diagrams with $W$ bosons (those with a $Z$ boson are relatively less
important), more precisely by special combinations of the non universal
components of vertices and boxes ("oblique" self-energies are only
contributing the RG component). Thus, at very high energy, the
relevance of "non-oblique" contributions appears essential, in
opposition to the situation that is met at the $Z$ peak.\par
Having computed the various gauge-invariant combinations, we shall then
evaluate in Section 3 the asymptotic expression of all the possible
observables (i.e. cross sections and asymmetries). Here the reliability
of the one loop expansion appears to depend critically on the c.m. energy
of the process. Typically, we find that, beyond a certain value that depends on
the chosen observable, the \underline{separate} gauge-invariant relative
effects begin to cross the dangerous ten percent value (orientatively
chosen as a realistic limit). This makes the validity of the
approximation rather doubtful if the requested precision must be below
the one percent level, and indicates the need of a proper resummation
already in the $TeV$ range.

As a byproduct of our calculation, we shall propose in Section 4 a simple
"effective"\\
 parametrization of all observables, aiming to provide a
satisfactory approximation of the one loop component 
in the energy range between $1~TeV$ and a
few $TeV$. We shall also compare this expression with the complete SM
calculations in the region below $1~TeV$ and discuss its possible
practical interest and its limitations in this not rigorously
asymptotic regime, in the particular theoretical 
framework provided by the
SM. A final discussion given in Section 5 will conclude the paper, and
Appendix A,B will contain the analytic expressions of the most relevant
one loop diagrams (self-energies, vertices and boxes) and the way they
contribute to the various observables.

\section{Asymptotic expressions of the gauge-invariant relevant
combinations at one loop.}

\subsection{Preliminary discussion}

The starting point of our paper, that has been already illustrated in
several previous references \cite{rising,Zsub}, is the choice of a
theoretical representation of four fermion neutral current processes,
strictly valid at one loop, that uses as theoretical input in addition
to the conventionally defined electric charge, the extremely high
precision measurements performed on top of the $Z$ resonance at
LEP1/SLC. The consequences of this attitude, in which the usual
parameter $G_{\mu}$ is replaced by $Z$ peak observables (this does not
introduce any appreciable theoretical error\cite{Zsub}), is that all
the physical information for the process $l^+l^-\to f\bar f$ (where $f$
is either a charged lepton, in which case $f\equiv l$, 
or a $u,d,s,c,b$ quark\footnote{Note that
our theoretical formalism cannot be applied in its present formulation
to the production of $t\bar t$ pairs for the simple reason that the
necessary LEP1/SLC information does not exist ($t\bar t$ cannot be
produced). The study of this process
requires modifications that are being studied.}, 
and \underline{for the moment} all external fermions are supposed to 
be massless) is provided by four different functions of $q^2$ and $\theta$
(the squared c.m. energy and scattering angle). The first one is the
"photonic form factor" ${\Delta}_{\alpha,
lf}(q^2,\theta)$ which is subtracted at $q^2=0$, whereas the three
other ones, $R_{lf}(q^2,\theta)$, $V^{\gamma Z}_{lf}(q^2,\theta)$ 
$V^{Z\gamma}_{lf}(q^2,\theta)$ corresponding to the one loop $ZZ$,
$\gamma Z$ and $Z\gamma$ transitions, are subtracted at $q^2=M^2_Z$.
All four functions are separately gauge independent, as discussed in
reference \cite{Zsub} on the basis of previous observations made by
Degrassi and Sirlin \cite{DS, DSpinch}. These functions receive two
types of one loop contributions: --- universal (i.e. $lf$-independent)
ones, coming from fermionic and bosonic self-energies as well as from a
part of some vertex diagrams; --- non-universal ones, coming from the
remaining parts of the vertex diagrams and from the box diagrams. As we
shall see in the next subsection, some universal contributions
correspond to the running of $\alpha$, $g^2_Z$ and $s^2_W$. The other
contributions lead asymptotically to the "Sudakov-type" terms that we
shall explicitely review in Appendix A. The way these four functions
enter the various $e^+e^- \to f\bar f$ observables will be shown in
Sect.3 and in Appendix B.

\subsection{RG contributions to the gauge-invariant combinations.}

Following our previous discussion, we shall first consider the
contributions to $\tilde{\Delta}_{\alpha}$, $R$, $V$ from the class of
diagrams that are supposed to reproduce the canonical RG asymptotic
behaviour. 
In the $R_\xi$ gauge in which we are working there will be no
contributions from the (finite) boxes to this sector. 
A simple request will select the possible contributions from
vertices. In fact, the various combinations that we consider are all,
as we stressed several times, gauge-independent. But the contribution
from self-energies is not such, owing to the set of bosonic "bubbles"
that must be considered. Therefore extra terms from vertices must be
properly added. Since self-energy contributions are of universal type,
the same property must obtain for the selected vertices. In practice,
this limits the choice to vertices where two $W$'s are involved. Note
that since we shall be working in the $\xi=1$ t'Hooft gauge, diagrams
with would-be Goldstone bosons must not be neglected in all
self-energies (in fact also, those with ghosts must be considered). In
the \underline{universal} component of the vertices, the
would-be contribution vanishes (in practice, it must only be retained in 
the not universal component of vertices with final 
$b\bar b$ pairs).\par
Our previous statement can be reformulated exactly using a previous
definition that can be found in Ref.\cite{DSpinch}. In fact, it would be
easy to show that the amount of $WW$ vertices that must be added to the
various self-energies is fully provided by the so-called "pinch"
component \cite{Cornwall}, and for more details we defer to that
reference. Otherwise stated, the combinations of self-energies and
"pinch" vertices that make up the RG behaviour in 
$\tilde{\Delta}_{\alpha}$, $R$, $V$ correspond rigorously to what was
called "gauge-invariant self-energies" by Degrassi and Sirlin
\cite{DS},\cite{DSpinch}.\par
In terms of Feynman diagrams, we must then consider the set represented
in Figs.(1),(2). The full expression of the various self-energies and
vertices of Figs.(1),(2) can be easily computed. One finds it in the
Appendix of Ref.(6), and we shall not rewrite it here since several
other new lengthy formulae will have to be written. From those
expressions one derives in a straightforward way the
\underline{universal} (i.e. $\theta,~lf$-independent) asymptotic
behaviour of $\tilde{\Delta}_{\alpha}$, $R$, $V$ that read, 
for $q^2>>\mu^2$:

\bq
\tilde{\Delta}^{(RG)}_{\alpha}(q^2,\theta)\to
{\alpha(\mu^2)\over12\pi}[{32\over3}N-21]ln({q^2\over \mu^2})
\label{daRG}\eq

\bq
R^{(RG)}(q^2,\theta)\to-{\alpha(\mu^2)\over4\pi
s^2c^2}[({20-40c^2+32c^4\over9}N+{1-2c^2-42c^4\over6}]
ln({q^2\over \mu^2})
\label{RRG}\eq

\bq
{c\over s}V^{(RG)}_{\gamma Z}(q^2,\theta)
={c\over s}V^{(RG)}_{Z\gamma}(q^2,\theta)
\to{\alpha(\mu^2)\over3\pi
s^2}[({10-16c^2\over6}N+{1+42c^2\over8}]
ln({q^2\over \mu^2})
\label{VRG}\eq
In the previous equations, $N$ is the number of fermion families,
$\mu$
is a "reference" scale that will be chosen following practical
arguments. In our special case, where a major part of the theoretical
input is fixed at the $Z$ mass, it seems rather natural to fix,
correspondingly, the value $\mu=M_Z$. This sets the scale at which the
asymptotic expression should become valid, $q^2>>M^2_Z$. 
Note that, in order to follow
consistently this choice, we should replace the theoretical input
$\alpha(0)$ with $\alpha(M^2_Z)$, even in the photonic component. This
does not introduce any substancial theoretical error, given the
available accuracy of the (theoretical) determination of
$\alpha(M^2_Z)$\cite{Jaeger}. Finally, choosing $\mu=M_Z$ suggests also to
identify the parameters $s^2$ , $c^2$ of eqs.(\ref{daRG}-\ref{VRG})
 with the experimentally
measured quantities $s_l^2(M_Z^2)$ and $c_l^2(M_Z^2)$ 
for which we shall take, following
the common attitude, the LEP1/SLD average \cite{expcombi}:

\bq
s_l^2(M_Z^2)=0.23157(18)
\label{expcombi}
\eq

To verify that eqs.(\ref{daRG}-\ref{VRG}) 
do indeed reproduce the running of the various Standard Model
couplings is now straightforward. The relevant expressions can be found
in previous references e.g. \cite{Marciano}. In order to make this
discussion reasonably self-consistent we write the two following
formulae:
\bq
\alpha^{(RG)}(q^2)=\alpha^{(RG)}(\mu^2)/\{1-[{\alpha^{(RG)}(\mu^2)\over
12\pi}({32\over3}N-21)]ln{q^2\over \mu^2}\}
\label{alRG}\eq
\bq
g^{2 (RG)}(q^2)=g^2(\mu^2)/\{1+[{g^2(\mu^2)\over
96\pi^2}(43-8N)]ln{q^2\over \mu^2}\}
\label{gRG}\eq
\noindent
where $\mu$ is the arbitrary reference scale. To derive the
corresponding expressions for $g^2_Z$, $s_l^2$ is immediate using
the corresponding definitions ($s_l^2\equiv{e^2\over g^2}$,
$g^2_Z\equiv {g^2\over1-{e^2\over g^2}}$).\par
One can thus realize that the RG running of $R(q^2,\theta)$ is that of
$g^2_Z$ and that of ${c\over s}V^{(RG)}_{\gamma Z}(q^2,\theta)$ 
is exactly that 
of $sin^2\theta_W$ as implied by the definitions of refs.\cite{Zsub, DS}
i.e. 
\bq
g^{(1)}_{Vl}(q^2, \theta)= I_{3L}-2Q_l s^2_l(q^2, \theta)
\eq

\noindent
with

\bq
\label{formal}
s^2_l(q^2, \theta) = 
\sin^2\theta_{W,0}+sc\tilde F_{\gamma Z, lf}(q^2, \theta) =
s^2_l(M^2_Z)[1+{c\over s}V^{\gamma Z}_{lf}(q^2,\theta)]
\eq
We have therefore derived the universal RG component of the gauge
invariant functions $\tilde{\Delta}_{\alpha}$, $R$, $V$ that leads to
an asymptotic (linear) logarithmic behaviour. From a formal point of
view, the existence of these components is due to the ultraviolet
\underline{divergence} of the generating Feynman diagrams (this is, of
course, cancelled in the physical subtracted combinations). In fact,
the coefficients of the logarithmic terms $ln q^2$ are exactly the same
(with opposite sign)
as those of the divergent terms $\propto 1/(d-4)$
in the various diagrams. When we turn to the task of evaluating other
possible asymptotic logarithmic contributions, we expect that these
will actually be produced by ultraviolet \underline{finite} quantities,
more precisely by non universal vertices and boxes, as it will be
discussed in the following Section (2.3).

\subsection{Sudakov-type contributions 
to the gauge-invariant combinations.}

{} From the previous discussion
we have learned that the set of one loop contributions that have not
been considered yet must be ultraviolet finite. This simple statement
already enables us to write the list of the remaining quantities, that
with our choice of $\xi=1$ gauge must be either vertices or boxes. More
precisely, we shall find here (a) the ``non pinch''
component of the vertices with two $W$s where the pure non RG behaviour
survives, Fig.~(2); (b) vertices with one $W$ or $Z$ exchange, 
Fig.~(3); and
(c) boxes, Fig.~(4).
Before performing the explicit
calculation of the asymptotic behaviour of the diagrams represented in
Figs.~(2-4), a brief preliminary discussion is now suited.

Infrared (IR) divergences arise in perturbative calculations 
from regions of integration
over the momentum $k$ where $k$ is small compared to the typical scales of
the process. 
This is a well known fact in
QED for instance, where  the problem of an
unphysical divergence is solved by giving the photon a fictitious mass
which acts as a cutoff for the IR divergent integral.
 When real (bremsstrahlung)
 and virtual contributions are summed, 
the dependence on this mass cancels and the
final result is finite.
 The (double)
logarithms   coming from these 
contributions are large and, 
  growing with the scale,
 can spoil perturbation theory 
and need to be resummed. They are usually 
called {\rm Sudakov } double logarithms \cite{Sudakov}.
 In the case of electroweak corrections,
similar logarithms arise when the typical scale of the process
 considered is much larger than the mass of the particles 
running in the loops, 
typically the $W(Z)$ mass \cite{Kuroda}. 
The expansion parameter results then
$\frac{\alpha}{4\pi s^2}ln^2\frac{q^2}{M_W^2}$, 
which is already
 10 \%  for for energies $\sqrt{q^2}$ 
of the order of $1~ TeV$.
This kind of corrections  becomes therefore
particularly relevant for next generation of 
linear colliders (LC \cite{muon}).
In the case of corrections coming from loops with $W(Z)$s, 
there is no  equivalent of ``bremsstrahlung'' like in
QED or QCD: the $W(Z)$, unlike the photon and the gluon,  
has a definite nonzero mass and
is experimentally detected like a separate particle. In this way the full
dependence on the $W(Z)$ mass is retained in the corrections.

In conclusion, for the process that  we consider here i.e. 
$l^+l^-\to f\bar{f}$ in the limit of
massless external fermions at the one loop level, we  expect
three kinds of contributions to become  ``large'' in the asymptotic
$q^2>>M_{W,Z}^2$ region:
\begin{itemize}
\item
Single logs: ($ln \frac{s}{\mu^2}$)  coming from U.V. 
divergences which can be reabsorbed by running of bare parameters.
\item
Single logs ($ln \frac{s}{m_{Z,W}^2}$)  coming from the analogue of QED
  collinear divergencies.
\item
Double logs ($ln^2 \frac{s}{m_{Z,W}^2}$)  coming from the analogue of QED
divergences that are of IR and collinear origin.

\end{itemize}

The double log   contributions come from 
vertex corrections  in which  one
gauge boson   is exchanged
 and from (direct and crossed) boxes  
with two $Z$s or two $W$s. 
The single collinear logs come also from  
vertex and box diagrams.
The single U.V. logs affect self energies and   
vertex.

Let us consider $\gamma \,(Z)\,\bar{f}\,f$ vertices first.
Using the definition 
\be
\Gamma_{\mu,f}^{\gamma\,(Z)}
\equiv
\bar{v}_f(p_1)\gamma_{\mu} (\V_{fL}^{\gamma\,(Z)} 
\,P_L+\V_{fR}^{\gamma\,(Z)} \,P_R)u_f(p_2)
\;\;\mbox{ with }\;\;
P_{L,R}=\frac{1\mp\gamma_5}{2}
\ee
we make an asymptotic expansion of the various vertices. 
Subtracting the 
``RG divergent component'' of the vertices that has been discussed 
previously, we obtain: 
\ba\label{inizio}
\V_{fL}^{\gamma}&=&i g\,s \, Q_f \,(
-~ \frac{1}{16 \pi^2}\, \frac{g^2}{c^2}\,
g_{fL}^{ 2} \,F[m_Z]
 -\\
&&\frac{1}{16 \pi^2}\, \frac{g^2}{2}\,
\frac{Q_{f'}}{Q_f}\,F[m_W]
-~\frac{1}{16 \pi^2} \,g^2\,
\frac{T_{3f}}{Q_f}\, G[m_W])
\\
 \V_{fR}^{\gamma}&=&ig \, s \,
 Q_f \,(
-~ \frac{1}{16 \pi^2}\, \frac{g^2}{c^2}\,
g_{fR}^{ 2} \,F[m_Z]
)
\ea
and
\ba
\V_{fL}^{Z}&=&i\frac{g}{c}\,  g_{fL}\,
 (- ~\frac{1}{16 \pi^2} \,\frac{g^2}{c^2}\,
g_{fL}^{ 2} \,
F[m_Z]
 -\\
&&
-~\frac{1}{16 \pi^2}\, \frac{g^2}{2}\,
\frac{g_{f'L}}{g_{fL}}\,
F[m_W]-\frac{1}{16 \pi^2}\, g^2\,
\frac{T_{3f} c^2}{g_{fL}}\, G[m_W] )
\\
\V_{fR}^{Z}&=&\frac{g}{c} \,
 g_{fR}\, (- ~\frac{1}{16 \pi^2}\, \frac{g^2}{c^2}\,
g_{fR}^{ 2} \,
F[m_Z])
\ea
where
\be
F[m]\equiv-4 \,ln\frac{q^2}{m^2}
+
ln^2\frac{q^2}{m^2};\;\;\;\;\;
G[m]\equiv-4 \, ln\frac{q^2}{m^2}
\ee
Here $f$  is the external fermion and $f'$ its isospin partner. 
Moreover, $g_{f(f')R}= -Q_{f(f')} s^2  $ and 
$ g_{f(f')L}=T_3^{f(f')}-Q_{f(f')}  s^2$
and $Q_f - Q_{f'}=2 T_{3f}$, $T_{3f}=-T_{3f'}$.

\vspace{0.3cm}

For the boxes, defining 
\ba \label{chiral}
\bar{v}_l(p_1)\gamma_\mu P_{L,R}u_l(p_2)\;
\bar{u}_f(p_3)\gamma_\mu P_{L,R}v_f(p_4)
\equiv \tilde{P}_{L,R}\otimes \tilde{P}_{L,R}
\ea
we have computed the corrections from direct and crossed box  diagrams 
 as a sum  of projected amplitudes on the left-right
 chiral basis: 
\ba \nonumber
 A^{Box}_{LL,lf}  \;\tilde{ P}_L \otimes \tilde{P}_L+
   A^{Box}_{LR,lf} \;  \tilde{P}_L \otimes \tilde{P}_R+\\ \nonumber
A^{Box}_{RL,lf} \; \tilde{P}_R \otimes \tilde{P}_L+
A^{Box}_{RR,lf} \;  \tilde{P}_R \otimes \tilde{P}_R
\ea
\noindent
(projecting on the "photon", "Z" Lorentz structures is then straightforward).

For the various components we find the following asymptotic expansions:
\ba
A^{Box}_{LL,lf}&=& \frac{\alpha}{4 \pi} \frac{e^2}{s^4} D_f^W
+\frac{\alpha}{\pi} \frac{e^2}{s^4c^4}
[g_{lL}^2 \; g_{fL}^2]
D^Z \\
A^{Box}_{LR,lf}&=&\frac{\alpha}{\pi} \frac{e^2}{s^4c^4}
[g_{lL}^2 \;  g_{fR}^2]
D^Z  \\
A^{Box}_{RL,lf}&=&\frac{\alpha}{\pi} \frac{e^2}{s^4c^4}
[g_{lR}^2 \;  g_{fL}^2]
D^Z  \\
A^{Box}_{RR,lf}&=& \frac{\alpha}{\pi} \frac{e^2}{s^4c^4}
[g_{lR}^2  \; g_{fR}^2]
D^Z
\ea
where  the functions that appear above are:

\ba \nonumber
D_{\mu,d}^W &= &
-\frac{1}{2q^2} ln^2\frac{q^2}{M_W^2}-\frac{1}{q^2}
\,ln \frac{1-\cos \theta}{2}  ln\frac{q^2}{M_W^2}
\\
D_{u}^W & = &\frac{1}{2\,q^2} \,ln^2\frac{q^2}{M_W^2}
+\frac{1}{q^2}  \, ln \frac{1+\cos \theta}{2}ln\frac{q^2}{M_W^2}
\ea
\ba\label{fine}
D^Z &=&
\frac{1}{q^2} \,ln \frac{1+\cos \theta}{1-\cos \theta} ln\frac{q^2}{M_Z^2} 
\ea
Eqns.(\ref{inizio}-\ref{fine}) are the main results of this paper. 
They contain in a compact form the leading asymptotic Sudakov-type 
(double and single log) contributions. 
Using the procedure given in \cite{Zsub} it is now 
straightforward to compute the corresponding contributions to
the four gauge-invariant combinations, that also depend on the chosen
final fermion. These are written in the complete form that is given 
in
Appendix A. For the practical purposes of our paper we shall now write
the numerical expressions of
$\tilde{\Delta}_{\alpha}$, $R$, $V$ that are obtained by summing the RG
contributions of eqs.(\ref{daRG}-\ref{VRG}) 
to those given in the Appendix A. Taking the value
of $s^2$ given in eq.(\ref{expcombi}) 
leads to the following result:\\

\bqa
\tilde{\Delta}_{\alpha,l\mu}(q^2,\theta)&\to&
{ \alpha(\mu^2)\over4\pi}\{
(-7.00+10.67
)\,ln{q^2\over\mu^2}+6.00\,ln^2{q^2\over M^2_W} 
\nonumber\\
&&+ 2.10\,ln{q^2\over M^2_Z}
-0.70\,ln^2{q^2\over M^2_Z}+\nonumber\\
&&
[ \,-2.00\,ln^2{q^2\over
M^2_W}-4.00\,ln{1-cos\theta\over2}ln{q^2\over M^2_W}+0.49\,
ln{1+cos\theta\over1+cos\theta}ln{q^2\over M^2_Z}
 \}\nonumber\\
&& \label{daasmu}\eqa
\bqa
R_{l\mu}(q^2,\theta)&\to&
{\alpha(\mu^2)\over4\pi}\{23.73-15.28
)\,ln{q^2\over\mu^2}-6.96\,ln{q^2\over M^2_W}-4.32\,ln^2{q^2\over M^2_W}
\nonumber\\
&&
-2.13\,ln{q^2\over M^2_Z}+0.71\,ln^2{q^2\over M^2_Z}+\nonumber\\
&&
[ \,6.64\,ln^2{q^2\over
M^2_W}+13.27\,ln{1-cos\theta\over2}ln{q^2\over M^2_W}-0.03\,
ln{1+cos\theta\over1+cos\theta}ln{q^2\over M^2_Z}]\}\nonumber\\
&&\label{Rasmu}
\eqa
\bqa
V_{\gamma Z,l\mu}(q^2,\theta)&=&V_{Z\gamma,l\mu}(q^2,\theta)\to
{\alpha(\mu^2)\over4\pi}\{(13.15-3.63
)\,ln{q^2\over\mu^2}-7.37\,ln{q^2\over M^2_W}-1.19\,ln^2{q^2\over M^2_W}
\nonumber\\
&&
-0.35\,ln{q^2\over M^2_Z}+0.12\,ln^2{q^2\over M^2_Z}+\nonumber\\
&&
[\, 3.64\,ln^2{q^2\over
M^2_W}+7.29\, ln{1-cos\theta\over2}ln{q^2\over M^2_W}-0.12\,
ln{1+cos\theta\over1+cos\theta}ln{q^2\over M^2_Z} ] \}\nonumber\\
&&\label{Vasmu}
\eqa

\bqa
\tilde{\Delta}_{\alpha,lu}(q^2,\theta)&\to&
{\alpha(\mu^2)\over4\pi}\{(-7.00+10.67
)\,ln{q^2\over\mu^2}+5.00\, ln{q^2\over M^2_W}
+0.33\,ln^2{q^2\over M^2_W}\nonumber\\
&&
 +1.95\,ln{q^2\over M^2_Z}
-0.65\,ln^2{q^2\over M^2_Z}+\nonumber\\
&&[-2.00\, ln^2{q^2\over
M^2_W}-4.00\, ln{1+cos\theta\over2}ln{q^2\over M^2_W}-0.63\,
ln{1+cos\theta\over1+cos\theta}ln{q^2\over M^2_Z}]\}
\nonumber\\
&& \label{daasu}\eqa
\bqa
R_{lu}(q^2,\theta)&\to&
{\alpha(\mu^2)\over4\pi}\{23.73-15.28
)\,ln{q^2\over\mu^2}-7.96\, ln{q^2\over M^2_W}-3.99\,ln^2{q^2\over M^2_W}
\nonumber\\
&&
-2.59\,ln{q^2\over M^2_Z}+0.86\,ln^2{q^2\over M^2_Z}
+\nonumber\\
&&[\,6.64\,ln^2{q^2\over
M^2_W}+13.27\,ln{1+cos\theta\over2}ln{q^2\over M^2_W}+0.16\,
ln{1+cos\theta\over1+cos\theta}ln{q^2\over
M^2_Z}] \}\label{Rasu}\nonumber\\
&&
\eqa
\bqa
V_{\gamma Z,lu}(q^2,\theta)&=&V_{Z\gamma,l\mu}(q^2,\theta)\to
{\alpha(\mu^2)\over4\pi}\{(13.15-3.63
)\,ln{q^2\over\mu^2}-5.55\, ln{q^2\over M^2_W}-1.79\,ln^2{q^2\over M^2_W}
\nonumber\\
&&
-1.00\,ln{q^2\over M^2_Z}+0.33\,ln^2{q^2\over M^2_Z}+\nonumber\\
&&
[\,3.64\,ln^2{q^2\over
M^2_W}+7.29\, ln{1+cos\theta\over2}ln{q^2\over M^2_W}+0.63\,
ln{1+cos\theta\over1+cos\theta}ln{q^2\over
M^2_Z}\}]\label{Vgzasu}\nonumber\\
&&
\eqa

\bqa
V_{Z\gamma,lu}(q^2,\theta)&\to&
{\alpha(\mu^2)\over4\pi}\{(13.15-3.63
)\,ln{q^2\over\mu^2}-7.92\,  ln{q^2\over M^2_W}-1.00\,ln^2{q^2\over M^2_W}
\nonumber\\
&&
-0.87\, ln{q^2\over M^2_Z}+0.29\,ln^2{q^2\over M^2_Z}+\nonumber\\
&&
[\,3.64\,ln^2{q^2\over
M^2_W}+7.29\,  ln{1+cos\theta\over2}ln{q^2\over M^2_W}+0.16\,
ln{1+cos\theta\over1+cos\theta}ln{q^2\over M^2_Z}] \}\nonumber\\
\label{Vzgasu}\eqa

\bqa
\tilde{\Delta}_{\alpha,ld}(q^2,\theta)&\to&
{\alpha(\mu^2)\over4\pi}\{(-7.00+10.67
)\,ln{q^2\over\mu^2}+4.00\, ln{q^2\over M^2_W}
+0.67\, ln^2{q^2\over M^2_W}\nonumber\\
&&
 +1.60\,ln{q^2\over M^2_Z}
-0.53\,ln^2{q^2\over M^2_Z}+\nonumber\\
&&[-2.00\, ln^2{q^2\over
M^2_W}-4.00\,ln{1-cos\theta\over2}ln{q^2\over M^2_W}+0.77\,
ln{1+cos\theta\over1+cos\theta}ln{q^2\over M^2_Z}]\}
\nonumber\\
&& \label{daasd}\eqa
\bqa
R_{ld}(q^2,\theta)&\to&
{\alpha(\mu^2)\over4\pi}\{(23.73-15.28
)\,ln{q^2\over\mu^2}-8.96\, ln{q^2\over M^2_W}-3.65\, ln^2{q^2\over M^2_W}
\nonumber\\
&&
-3.64\,ln{q^2\over M^2_Z}+1.21\,ln^2{q^2\over M^2_Z}+\nonumber\\
&&
[\,6.64\,ln^2{q^2\over
M^2_W}+13.27\,ln{1-cos\theta\over2}ln{q^2\over M^2_W}-0.29\,
ln{1+cos\theta\over1+cos\theta}ln{q^2\over
M^2_Z}] \}\label{Rasd}\nonumber\\
&&
\eqa
\bqa
V_{\gamma Z,ld}(q^2,\theta)&\to&
{\alpha(\mu^2)\over4\pi}\{(13.15-3.63
)\,ln{q^2\over\mu^2}-3.73\, ln{q^2\over M^2_W}-2.40\,ln{q^2\over M^2_W}
\nonumber\\
&&
-0.91\,ln{q^2\over M^2_Z}+0.30\,ln^2{q^2\over M^2_Z}+\nonumber\\
&&
[\,3.64\,ln{q^2\over
M^2_W}+7.29\, ln{1-cos\theta\over2}ln{q^2\over M^2_W}-1.15\,
ln{1+cos\theta\over1+cos\theta}ln{q^2\over
M^2_Z}]   \}\label{Vgzasd}\nonumber\\
&&
\eqa

\bqa
V_{Z\gamma,ld}(q^2,\theta)&\to&
{\alpha(\mu^2)\over4\pi}\{(13.15-3.63
)\,ln{q^2\over\mu^2}-8.47\,  ln{q^2\over M^2_W}-0.82\,ln^2{q^2\over M^2_W}
\nonumber\\
&&
-1.62\, ln{q^2\over M^2_Z}+ 0.54\,ln^2{q^2\over M^2_Z}+\nonumber\\
&&
[\,3.64\,ln^2{q^2\over
M^2_W}+7.29\, ln{1-cos\theta\over2}ln{q^2\over M^2_W}-0.19\,
ln{1+cos\theta\over1+cos\theta}ln{q^2\over M^2_Z}]\}\nonumber\\
\label{Vzgasd}\eqa
We have left three scale parameteres $\mu,M_W,M_Z$ to evidentiate the
different origin (RGEs, W diagrams, Z diagrams) of the various
contributions.
In the first two lines of each 
expression
 there are the contributions coming from bubbles and vertices.
In the RGE part of them, the contributions of fermionic and
bosonic degrees of freedom are left separated.
In practice for numerical estimates we set
$\mu=M_Z=91.187$ GeV.
Finally, the contributions from the $WW$ and
$ZZ$ box
diagrams are systematically grupped in the last three terms inside 
the square brackets.

Eqs.(\ref{daasmu}-\ref{Vzgasd}) allow us 
to derive the leading asymptotic behaviour of all the relevant
observables (i.e. cross sections and asymmetries) of the considered
four-fermion process. This will be done in the following Section.

\section{Asymptotic expression of the physical observables}

Starting from eqs.(\ref{daasmu}-\ref{Vzgasd}) 
it is now straightforward to derive the asymptotic
behaviour of any given observable. In order to proceed in a systematic
way, we shall start from the very general formula:

\bq
{d\sigma_{lf}\over dcos\theta}={4\pi\over3}\N_fq^2\{{3\over8}
(1+cos^2\theta)[(1-PP')U_{11}+(P'-P)U_{21}]+{3\over4}cos\theta[(1-PP')
U_{12}+(P'-P)U_{22}]\}
\label{sig}\eq
\noindent
where the quantities $U_{ij}$ are defined in terms of
$\tilde{\Delta}^{(lf)}$, $R^{(lf)}$, $V^{(lf)}_{\gamma
Z}$, $V^{(lf)}_{Z\gamma}$
in Appendix B, 
$P,P'$ are the (conventionally defined) \underline{longitudinal} 
polarization degree of
the initial lepton and antilepton, and $\N_f$ is 
the colour factor for the $f\bar f$ channel which includes the
appropriate QCD corrections to the input.

{} From the previous equation, and from eqs.(\ref{daasmu}-\ref{Vzgasd}), 
we can now derive the
asymptotic expansion of the relevant observables. We shall first 
consider the case of unpolarized
beams and concentrate our attention on a set of "typical" observables
that one expects to measure at future colliders. These will be: the
cross section for muon (and/or tau) production $\sigma_{\mu}$, the
related forward-backward asymmetry $A_{FB,\mu}$ and  the cross section for
"light" $u,d,s,c,b$ production $\sigma_5$.

As we anticipated in the Introduction, we shall treat quark
production in the massless quark limit. 
This will introduce a certain approximation
in the treatment of the cross section for $b$ production $\sigma_b$ where
the effects of the non vanishing top mass in the asymptotic regime must be 
carefully estimated. Since the contribution of $\sigma_b$ to $\sigma_5$ is
relatively small, we shall not treat this extra effect in this paper. We 
shall rather postpone a complete rigorous discussion of $m_t$ effects
to a forthcoming article.

Collecting all our numerical formulae leads to the final
asymptotic expressions of the previous observables, that read
($N=3$ is the number of fermion families and $\mu^2=M^2_Z\approx M^2_W$):
\bqa \label{sigm}
\sigma_{\mu}&=&\sigma^{B}_{\mu}
[1+{\alpha\over4\pi}
\{(7.72\,N-20.58
)\,ln{q^2\over\mu^2}+35.27\, ln{q^2\over M^2_W}-4.59\, ln^2{q^2\over M^2_W}
\nonumber\\
&&
+4.79\,  ln{q^2\over M^2_Z} -1.43\, ln^2{q^2\over M^2_Z}+........\}]
\label{sigmu}\eqa
\bqa\label{ass}
A_{FB,\mu}&=&A^{B}_{FB,\mu}+{\alpha\over4\pi}\{(0.54\,N-5.90
)\,ln{q^2\over\mu^2}+10.19\, ln{q^2\over M^2_W}-0.08\,ln^2{q^2\over M^2_W}
\nonumber\\
&&
+1.25\, ln{q^2\over M^2_Z} -0.004\,ln^2{q^2\over M^2_Z}+........\}
\label{AFBmu}\eqa
\bqa\label{sigm5}
\sigma_{5}&=&
\sigma^{B}_{5}[1+ { \alpha\over4\pi}
\{ (9.88\,N-42.66
)\, ln{q^2\over\mu^2}+46.58\, ln{q^2\over M^2_W}-6.30\, ln^2{q^2\over M^2_W}
\nonumber\\
&&
+7.25\,ln{q^2\over M^2_Z} -2.03\,ln^2{q^2\over M^2_Z}+........\}]
\nonumber\\
\eqa

The precise definition of the ``Born'' (B) quantities that 
appear in these equations will be given in the forthcoming section.
The dots that appear in the brackets  
refer to the "non-leading" terms. These could 
either be constants or $O({1\over q^2})$ components 
whose asymptotic effect vanishes. At non asymptotic 
energies, such terms (in particular the constants) 
should be discussed; we shall return on this point in 
the following Section.

Next, we have considered the simplest case of full lepton longitudinal
polarization and treated one simple observable: the
longitudinal polarization asymmetry for final lepton production,
$A_{LR,\mu}$. 
Using the same procedure leads to the following expressions

\bqa
A_{LR,\mu}&=&A^{B}_{LR,\mu}+{\alpha\over4\pi}\{(1.82\,N-19.79
)\,ln{q^2\over\mu^2}+30.76\, ln{q^2\over M^2_W}-3.52\, ln^2{q^2\over M^2_W}
\nonumber\\
&&
+0.78\,ln{q^2\over M^2_Z} -0.17\,ln^2{q^2\over M^2_Z}+.......\}.
\label{ALRmu}\eqa
where the ``Born'' asymmetry will be defined later on.

Eqs.(\ref{sigmu}-\ref{ALRmu}) 
are the physical predictions of this paper. Looking at their
expressions one immediately notices the following main features:\\
a) The coefficient of the RG linear logarithm is much smaller than that
of the Sudakov one. A naive expectation of an asymptotic behaviour
essentially reproduced by the RG logarithms would be therefore, at this stage, 
completely wrong. In other words, the high energy behaviour of SM
observables is only partially reabsorbed by the running of the coupling
constants.\\
b) The role of the Sudakov squared logarithm is numerically relevant in
almost all the considered observables, with the exception of the muon
asymmetry where it almost vanishes.\\
c) Both the linear and the quadratic logarithmic terms are, separately
taken, relatively "large". However, at the considered one loop level,
they have opposite sign and the overall contribution is smaller. This
raises a few important questions that we shall try now to
investigate.\par
The first question is that of whether the various logarithmic terms
have separately a physical meaning. In our opinion, at the considered
one loop level, this must be evidently the case. In fact, since the
overall expansion is necessarily gauge-independent, the different
powers of the logarithm must satisfy the same request (note that this
applies, in general, to the \underline{sum} of the separate
contributions from vertices and boxes). A priori, one expects that a
gauge-independent quantity might (should) have a physical meaning. For
the RG component, this is connected with the running of the couplings.
For the remaining terms, their origin is related, as we have seen previously,
to the fact that a final state $W$ is supposed to be experimentally detectable. 

The second question is that of the reliability of 
the one loop perturturbative electroweak corrections to the considered 
observables. 
This is fixed by  the aimed theoretical accuracy, which in turn is
dictated by the expected experimental precision. For the future $l^+l^-$
colliders we shall stick to a conservative expectation of about 1 \%
experimental  (relative) accuracy.
In this spirit,we shall assume a "reliability bareer" for 
one loop effects         
at the relative 10\% level; of course, these numbers can be easily changed 
without modifying  
the philosophy of the approach.
Having fixed the accuracy level, we can proceed following 
two different
criteria. The first one is a global one: if the total 
$O(\alpha(\mu)/\pi)$
correction remains below the ten percent threshold, we can consider
the perturbative  expansion to be under control and the 
one loop approximation
to be a satisfactory one. The second point of view,
the most correct in our opinion, is that in which one
requires that \underline{all} the different logarhitmic 
effects individually
satisfy the previous 10 \% criterium. Starting from these 
requirements, we
have therefore examined how the relative size of the various components in the
considered observables varies with energy, in a energy range where our
asymptotic expansion might be reasonably accurate. 
Naively we would expect that this corresponds orientatively to energies of
about 1 TeV
and beyond. 
In fact, we have verified by an exact one loop numerical calculation
of the various observables, where all the contributions from the various
diagrams are retained without approximation, that the rigorous expressions
are reproduced by our asymptotic expansions within a few percent at most,
in the energy range between 1 TeV and 10 TeV (larger energy values seem to 
us not realistic). We can therefore reasonably conclude that, in this energy range,
Eqs.~(\ref{sigm}--\ref{sigm5}) contain the bulk of the one loop electroweak corrections,
and can be used for a meaningful discussion of the size of the various
effects.

The results of our analysis are shown in Figs. 5-8.
As one sees from inspection of these figures, the situation is quite different
for the two criteria and for the various observables. To be more precise, we
list the various cases separately:

a) $\sigma_\mu$. Here the global relative effect remains below the 10 \% limit
 in the full range $1~ TeV < \sqrt{q^2} < 10~ TeV$. 
However, the individual
 Sudakov components both cross the ``safety limit'' practically in the full
 range.

b) $\sigma_5$. Here the global relative effect rises beyond the ``safety
 limit'' at energies larger than $4~ TeV$. The separate Sudakov contributions are
 over the limit in the full considered range.

c) $A_{FB,\mu}$. Here the global relative effects remain always below the
limit. The linear Sudakov crosses the limit at about $3~ TeV$ (the quadratic one
is almost absent in this case)
  
d)  $A_{LR,\mu}$. In principle, both global and individual relative effects are
systematically well beyond the ten percent limit in the full range. A word of
caution is, though, suited since for this special observable the Born
approximation is particularly small ($\approx 0.07$). Moreover,
the most important one loop effects 
are produced by $W$ diagrams, that only contribute to the lefthanded cross
section and generate therefore particularly large effects in this observable.

The conclusion of this preliminary investigation is that, in the considered
energy range beyond $1~ TeV$, the validity of the one loop SM electroweak
expansion is debatable. The necessity of a two loop calculation, leading if
possible to a resummation af Sudakov effects, appears to us strongly
motivated.

Having examined the situation that occurs in the region beyond $1~ TeV$, we want
now to turn our attention to the energy region below. Here an asymptotic
expansion is certainly less justified. We shall discuss the emerging picture
in next Section.

\section{A simple formula for cross sections and asymmetries below 1 TeV}

In the previous sections we have derived a one loop asymptotic expansion for
various observables and we have explicitly shown in Section 3
that in the energy range beyond $1~ TeV$, where the expansion is able to
describe well the one loop corrections, the reliability of the one loop 
approximation
is not evident. These conclusions would be particularly relevant for a
possible future muon collider, operating in a energy region of a few
$TeV$. On
the other hand, the simplicity and the well established physical
interpretation of the various terms of the expansion pushes us to investigate 
the possibility
of using them in the energy range \underline{below} $1~ TeV$, where,
in principle, the validity of an asymptotic expansion is not guaranteed. In
particular, we are interested in simple formulas that describe cross sections
and asymmetries e.g. in the energy range $300~ GeV < \sqrt{q^2} < 1~ TeV$, 
where the
future linear colliders (LC) will be operating at an energy that we assume to
be of $500~ GeV$. For what concerns the experiment accuracy, we shall stick to
the previous conservative assumption of a one percent precision, which is 
rather pessimistic  compared to the latest expectations \cite{LC}.

Our empirical procedure has been the following. We have first plotted the
\underline{exact} SM predictions for $\sigma_\mu$ and $A_{FB, \mu}$ in the 
considered energy region
(a full discussion of $\sigma_5$, that requires a rigorous calculation
of the top effects will be given in the annunciated following paper). 
The calculation has been made using the program
TOPAZ0~\cite{LEP2}\footnote{
We thank G. Passarino for kindly providing us with the numerical data.
}. We have then compared the TOPAZ0 
one loop results with those of the simple asymptotic
expansion eqns.~(\ref{sigm}-\ref{sigm5}). A priori, we expect that ``some'' modifications of
our formulae are unavoidably requested. 

The results of our first comparison are shown in Fig.~9. As one sees,
the difference between the full calculation and the logarithmic approximation
remains impressively practically constant in the whole considered energy range. This (in our
opinion) remarkable fact allows to conclude that the simplest modification of the logarithmic
expansion obtained by adding a constant term, seems sufficient to obtain a very accurate 
agreement.

For the two observables we therefore write
\be
\label{simple1}
\sigma_{\mu} = \sigma^B_{\mu} ( 1 + \frac{\alpha}{4\pi}(c_{\mu} + \mbox{logarithms}))
\ee
\be
\label{simple2}
A_{FB} = A_{FB}^B + \frac{\alpha}{4\pi}(c_{FB} + \mbox{logarithms})
\ee
where, "logarithms" stands for the logarithmic terms of eqs.(\ref{sigm}-
\ref{sigm5}) and where
in agreement with our overall philosophy, $\sigma_\mu^B$ and  $A_{FB, \mu}^B$
are the Born level expressions in which
we have set $\alpha=\alpha(M_Z^2)$. They are obtained , in a
straightforward way from the expressions given in eq.(\ref{U11pro}
-\ref{U22pro}), putting all 1-loop terms  
$\tilde{\Delta}_{\alpha,lf}$, $R_{lf}$ , $V_{\gamma Z,lf}$,
$V_{Z\gamma,lf}$ equal to zero and replacing $\alpha(0)$ by
$\alpha(M_Z^2)$.
In the case of the leptonic cross section we used
\begin{eqnarray}
\label{bornmu}
\lefteqn{\sigma^B_\mu = \sigma^{B, \gamma\gamma}_\mu +\sigma^{B, 
\gamma Z}_\mu +\sigma^{B, ZZ}_\mu = \frac{4\pi \alpha^2(M_Z^2)}{3q^2} +} && \\
&+& 8\pi\alpha(M_Z^2)\ \frac{\Gamma_\mu}{M_Z}\  
\frac{q^2-M_Z^2}{(q^2-M_Z^2)^2+\Gamma_Z^2 M_Z^2}\
\frac{\tilde{v}_l^2}{1+\tilde{v}_l^2}
+ 12\pi \frac{\Gamma_\mu^2}{M_Z^2}\frac{q^2}
{(q^2-M_Z^2)^2+\Gamma_Z^2 M_Z^2} \nonumber
\end{eqnarray}
with $\tilde{v}_{l} =1-4 s_l^2(M_Z^2)$.
For the forward-backward asymmetry we write

\bq
A^B_{FB,\mu}={\sigma_{FB,\mu}^B\over\sigma^B_\mu}
\eq
\noindent
with
\begin{eqnarray}
\label{bornass}
\lefteqn{\sigma_{FB,\mu}^B = \sigma_{FB,\mu}^{B, \gamma Z} 
+\sigma_{FB,\mu}^{B, ZZ} = } && \\
&=& 6\pi\alpha(M_Z^2)\ \frac{\Gamma_\mu}{M_Z}\  
\frac{q^2-M_Z^2}{(q^2-M_Z^2)^2+\Gamma_Z^2 M_Z^2}\
\frac{1}{1+\tilde{v}_l^2}
+ 36\pi \frac{\Gamma_\mu^2}{M_Z^2}\frac{q^2}
{(q^2-M_Z^2)^2+\Gamma_Z^2 M_Z^2}\ \frac{\tilde{v}^2}{(1+\tilde{v}^2)^2} \nonumber
\end{eqnarray}
The choice of the constants is, to some extent, arbitrary. We have 
followed the rather pragmatic attitude of optimizing our approximation
in the 500 GeV region (LC domain). This fixes the values~\cite{data}:
\begin{equation}
c_\mu = -53.87, \qquad
c_{FB} = -20.82
\end{equation}

The comparison of our simple effective expressions 
Eqs.~(\ref{sigm}-\ref{bornass}) 
with TOPAZ0 are shown again in Fig.~9. As one sees,
between $300~GeV$ and $1~TeV$, 
the difference between the two calculations is at the permille level for both the 
considered observables.
Therefore, our approximate expressions could be safely used to
perform, e.g. preliminary analyses of possible effects of
competitor electroweak models~\footnote{
The simplest example would be provided by a class of models with anomalous
triple gauge couplings (AGC)~\cite{hagi}. Here, the effect of the models is that of
introducing linear terms of the type $q^2/\Lambda^2$ ~\cite{AGCus}, 
thus changing the energy
dependence of the observables as a function of the AGC parameters. 
Another application could be one that 
requires integration over a certain fraction of the energy range
involving the product of the cross section by known weight functions, like those encountered when computing QED
convolution. This could be performed in a rather economical way.}.

As a final comment, we would like to notice that the fact that an asymptotic
formula well describes the behavior of the exact corrections in a (relatively) 
low energy
range can be due to the the absence of structure like resonances in that
region( in our analysis we have ignored the small peak at the $t\bar{t}$
production threshold which is negligible in any case for the considered
observables).

\section{Conclusions}
 In this paper we have determined the energy growing electroweak logarithms
of RG(UV) and Sudakov(IR) origin that appear in the asymptotic expansions of
four (massless) fermions processes. We have shown that (linear and quadratic)
Sudakov-type logarithms are numerically much more important than RG-driven
logarithms in the TeV region. Moreover, although the overall effect on the
observables are of the order of a few percent, the single contributions are of
the 10\% order in this region. This supports the idea that next order
calculations and/or resummations are needed. On the other hand, our simple
one loop expansions  might reveal to be unexpectedly useful in the energy
region \underline{below} 
1 TeV, since the correct energy dependence of the observables
seems to be properly described. 

The fact that a satisfactory description of electroweak corrections at
asymptotic energies requires more complicated descriptions beyond the
``naive'' one loop approximation would not be, in our opinion, a surprising
feature for a gauge theory. For QED and QCD similar problems of higher order
calculations and resummation have been notoriously already preformed. Our
personal impression is that a proper description of the genuine electroweak
sector of the SM at the future relevant energies in the $TeV$ region requires an
analogous remarkable effort.

\newpage

\renewcommand{\theequation}{A.\arabic{equation}}
\renewcommand{\thesection}{A.\arabic{section}}
\setcounter{equation}{0}
\setcounter{section}{0}

{\large \bf Appendix A: Sudakov-type contributions}\\

{\bf  Final fermions $f\neq b$}

\bqa
\tilde{\Delta}^{(S)}_{\alpha,lf}(q^2,\theta)&\to&
{\alpha\over4\pi}[6-\delta_u-2\delta_d]ln{q^2\over M^2_W}
+{\alpha\over12\pi}(\delta_u+2\delta_d)ln^2{q^2\over M^2_W}
+{\alpha(2-v^2_l-v^2_f)\over64\pi s^2c^2}[3ln{q^2\over M^2_Z}
-ln^2{q^2\over M^2_Z}]\nonumber\\
&&-{\alpha\over2\pi}[(ln^2{q^2\over M^2_W}+2ln{q^2\over
M^2_W}ln{1-cos\theta\over2})(\delta_{\mu}+\delta_d)
+(ln^2{q^2\over M^2_W}+2ln{q^2\over
M^2_W}ln{1+cos\theta\over2}))\delta_u]\nonumber\\
&&-{\alpha\over256\pi Q_f
s^4c^4}[(1-v^2_l)(1-v^2_f)(ln{q^2\over
M^2_Z}ln{1+cos\theta\over1+cos\theta})]\nonumber\\&&
\eqa

\bqa
R^{(S)}_{lf}(q^2,\theta)&\to&
-{3\alpha\over4\pi s^2}[2c^2-\delta_{\mu}-(1-{s^2\over3})\delta_u
-(1-{2s^2\over3})\delta_d]ln{q^2\over M^2_W}\nonumber\\
&&
-{\alpha\over4\pi s^2}
[\delta_{\mu}+(1-{s^2\over3})\delta_u+(1-{2s^2\over3})\delta_d]
ln^2{q^2\over M^2_W}\nonumber
\\
&& 
-{\alpha(2+3v^2_l+3v^2_f)\over64\pi s^2c^2}[3ln{q^2\over M^2_Z}
-ln^2{q^2\over M^2_Z}]\nonumber\\
&&
+{\alpha c^2\over2\pi s^2}[(ln^2{q^2\over M^2_W}+2ln{q^2\over
M^2_W}ln{1-cos\theta\over2})(\delta_{\mu}+\delta_d)
+(ln^2{q^2\over M^2_W}+2ln{q^2\over
M^2_W}ln{1+cos\theta\over2}))\delta_u]\nonumber\\
&&
+I_{3f}{\alpha\over2\pi s^2c^2}[v_lv_fln{q^2\over
M^2_Z}ln{1+cos\theta\over1+cos\theta}]\nonumber\\&&
\eqa

\bqa
V^{(S)}_{\gamma Z,lf}(q^2,\theta)&\to&
{\alpha \over8\pi cs}([3-12c^2+2c^2(\delta_u+2\delta_d)]
ln{q^2\over M^2_W}-[1+{2\over3}c^2(\delta_u+2\delta_d)]
ln^2{q^2\over M^2_W})\nonumber\\
&&
-[{\alpha v_l(1-v^2_l)\over128\pi s^3c^3}+{\alpha|Q_f|v_f\over8\pi sc}]
[3ln{q^2\over M^2_Z}-ln^2{q^2\over M^2_Z}]\nonumber\\
&&
+{\alpha c\over2\pi s}
[(ln^2{q^2\over M^2_W}+2ln{q^2\over
M^2_W}ln{1-cos\theta\over2})(\delta_{\mu}+\delta_d)
+(ln^2{q^2\over M^2_W}+2ln{q^2\over
M^2_W}ln{1+cos\theta\over2}))\delta_u]\nonumber\\
&&
+I_{3f}{\alpha\over16\pi s^3c^3}[v_f(1-v^2_l)ln{q^2\over
M^2_Z}ln{1+cos\theta\over1+cos\theta}]\nonumber\\&&
\eqa

\newpage

\bqa
V^{(S)}_{Z\gamma,lf}(q^2,\theta)&\to&
{\alpha \over8\pi cs}([3-12c^2-2s^2(\delta_u+2\delta_d)]
ln{q^2\over M^2_W}-[1-{2\over3}s^2(\delta_u+2\delta_d)]
ln^2{q^2\over M^2_W})\nonumber\\
&&
-[{\alpha v_f(1-v^2_f)\over128\pi |Q_f|s^3c^3}
+{\alpha v_l\over8\pi sc}]
[3ln{q^2\over M^2_Z}
-ln^2{q^2\over M^2_Z}]\nonumber\\
&&
+{\alpha c\over2\pi s}
[(ln^2{q^2\over M^2_W}+2ln{q^2\over
M^2_W}ln{1-cos\theta\over2})(\delta_{\mu}+\delta_d)
+(ln^2{q^2\over M^2_W}+2ln{q^2\over
M^2_W}ln{1+cos\theta\over2}))\delta_u]\nonumber\\
&&
+{\alpha\over32\pi Q_f s^3c^3}[v_l(1-v^2_f)ln{q^2\over
M^2_Z}ln{1+cos\theta\over1+cos\theta}]\nonumber\\&&
\eqa

\noindent
where $\delta_{\mu,u,d}=1$ for $f=\mu,u,d$ and 0 otherwise and
$v_l=1-4 \,s^2,\;\;v_f=1-4 \,|Q_f|\,s^2 $.\par
In each of the above equations, we have successively added the
contributions coming from triangles containing one or two $W$,
triangles containing one $Z$, from $WW$ box and finally from $ZZ$
box.

\newpage

\renewcommand{\theequation}{B.\arabic{equation}}
\renewcommand{\thesection}{B.\arabic{section}}
\setcounter{equation}{0}
\setcounter{section}{0}

{\large \bf Appendix B: Contributions to the various observables}\\

The general expression of the $l^+l^-\to f\bar f$ cross section
can be written as

\bq
{d\sigma_{lf}\over dcos\theta}={4\pi\over3}\N_fq^2\{{3\over8}
(1+cos^2\theta)[(1-PP')U_{11}+(P'-P)U_{21}]+{3\over4}cos\theta[(1-PP')
U_{12}+(P'-P)U_{22}]\}
\label{sigA}\eq
\noindent
where
\bqa  
U_{11}=&&
{\alpha^2(0)Q^2_f\over q^4}[1+2\tilde{\Delta}^{(lf)}\alpha(q^2,\theta)]
\nonumber\\
&&+2[{\alpha(0)|Q_f|}]{q^2-M^2_Z\over
q^2((q^2-M^2_Z)^2+M^2_Z\Gamma^2_Z)}[{3\Gamma_l\over
M_Z}]^{1/2}[{3\Gamma_f\over N_f M_Z}]^{1/2}
{\tilde{v}_l \tilde{v}_f\over
(1+\tilde{v}^2_l)^{1/2}(1+\tilde{v}^2_f)^{1/2}}\nonumber\\
&&\times[1+
\tilde{\Delta}^{(lf)}\alpha(q^2,\theta) -R^{(lf)}(q^2,\theta)
-4s_lc_l
\{{1\over \tilde{v}_l}V^{(lf)}_{\gamma Z}(q^2,\theta)+{|Q_f|\over\tilde{v}_f}
V^{(lf)}_{Z\gamma}(q^2,\theta)\}]\nonumber\\ 
&&+{[{3\Gamma_l\over
M_Z}][{3\Gamma_f\over \N_f M_Z}]\over(q^2-M^2_Z)^2+M^2_Z\Gamma^2_Z}
\nonumber\\
&&\times[1-2R^{(lf)}(q^2,\theta)
-8s_lc_l\{{\tilde{v}_l\over1+\tilde{v}^2_l}V^{(lf)}_{\gamma
Z}(q^2,\theta)+{\tilde{v}_f|Q_f|\over(1+\tilde{v}^2_f)}
V^{(lf)}_{Z\gamma}(q^2,\theta)\}]
\label{U11pro}
\eqa
\bqa
U_{12}=&& 2[{\alpha(0)|Q_f|}]{q^2-M^2_Z\over
q^2((q^2-M^2_Z)^2+M^2_Z\Gamma^2_Z)}
[{3\Gamma_l\over M_Z}]^{1/2}[{3\Gamma_f\over \N_f
M_Z}]^{1/2}{1\over(1+\tilde{v}^2_l)^{1/2}(1+\tilde{v}^2_f)^{1/2}}
\nonumber\\
&&\times[1+
\tilde{\Delta}^{(lf)}\alpha(q^2,\theta) -R^{(lf)}(q^2,\theta)]\nonumber\\
&&+{[{3\Gamma_l\over M_Z}][{3\Gamma_f\over \N_f
M_Z}]\over(q^2-M^2_Z)^2+M^2_Z\Gamma^2_Z}
[{4\tilde{v}_l \tilde{v}_f\over(1+\tilde{v}^2_l)(1+\tilde{v}^2_f)}]
\nonumber\\
&&\times[1-2R^{(lf)}(q^2,\theta)-4s_lc_l
\{{1\over \tilde{v}_l}V^{(lf)}_{\gamma Z}(q^2,\theta)+{|Q_f|\over\tilde{v}_f}
V^{(lf)}_{Z\gamma}(q^2,\theta)\}]
\label{U12pro}
\eqa
\bqa
U_{21}=&& 2[{\alpha(0)|Q_f|}]{q^2-M^2_Z\over
q^2((q^2-M^2_Z)^2+M^2_Z\Gamma^2_Z)}
[{3\Gamma_l\over
M_Z}]^{1/2}[{3\Gamma_f\over \N_f
M_Z}]^{1/2}{\tilde{v}_f\over(1+\tilde{v}^2_l)^{1/2}
(1+\tilde{v}^2_f)^{1/2}}\nonumber\\
&&\times[1+\tilde{\Delta}^{(lf)}\alpha(q^2,\theta) -R^{(lf)}(q^2,\theta)
-{4s_lc_l|Q_f|\over\tilde{v}_f}V^{(lf)}_{Z\gamma}(q^2,\theta)]\nonumber\\
&&+{[{3\Gamma_l\over
M_Z}][{3\Gamma_f\over \N_f
M_Z}]\over(q^2-M^2_Z)^2+M^2_Z\Gamma^2_Z}
[{2\tilde{v}_l \over(1+\tilde{v}^2_l)}]\nonumber\\
&&\times[1-2R^{(lf)}(q^2,\theta)-4s_lc_l
\{{1\over \tilde{v}_l}V^{(lf)}_{\gamma
Z}(q^2,\theta)+{2\tilde{v}_f|Q_f|\over(1+\tilde{v}^2_f)}
V^{(lf)}_{Z\gamma}(q^2,\theta)\}]   
\label{U21pro}
\eqa
\bqa
U_{22}= && 2[{\alpha(0)|Q_f|}]{q^2-M^2_Z\over
q^2((q^2-M^2_Z)^2+M^2_Z\Gamma^2_Z)}
[{3\Gamma_l\over
M_Z}]^{1/2}[{3\Gamma_f\over \N_f
M_Z}]^{1/2}{\tilde{v}_l\over(1+\tilde{v}^2_l)^{1/2}
(1+\tilde{v}^2_f)^{1/2}}\nonumber\\
&&\times[1+\tilde{\Delta}^{(lf)}\alpha(q^2,\theta) -R^{(lf)}(q^2,\theta)
-{4s_lc_l\over \tilde{v}_l}V^{(lf)}_{\gamma Z}(q^2)]\nonumber\\
&&+{[{3\Gamma_l\over
M_Z}][{3\Gamma_f\over \N_f
M_Z}]\over(q^2-M^2_Z)^2+M^2_Z\Gamma^2_Z}
[{2\tilde{v}_f \over(1+\tilde{v}^2_f)}]\nonumber\\
&&\times[1-2R^{(lf)}(q^2,\theta)-4s_lc_l
\{{2\tilde{v}_l\over(1+\tilde{v}^2_l)}
V^{(lf)}_{\gamma Z}(q^2,\theta)+{|Q_f|\over
\tilde{v}_f}V^{(lf)}_{Z\gamma}(q^2,\theta)\}] \label{U22pro}
\eqa 

Here $P,P'$ are the \underline{longitudinal} 
polarization degree of
the initial lepton and antilepton, and $\N_f$ is 
the colour factor for the $f\bar f$ channel which includes the
appropriate QCD corrections to the input.

\begin{figure}[p]
\[
\hspace{-2.5cm}
\epsfig{file=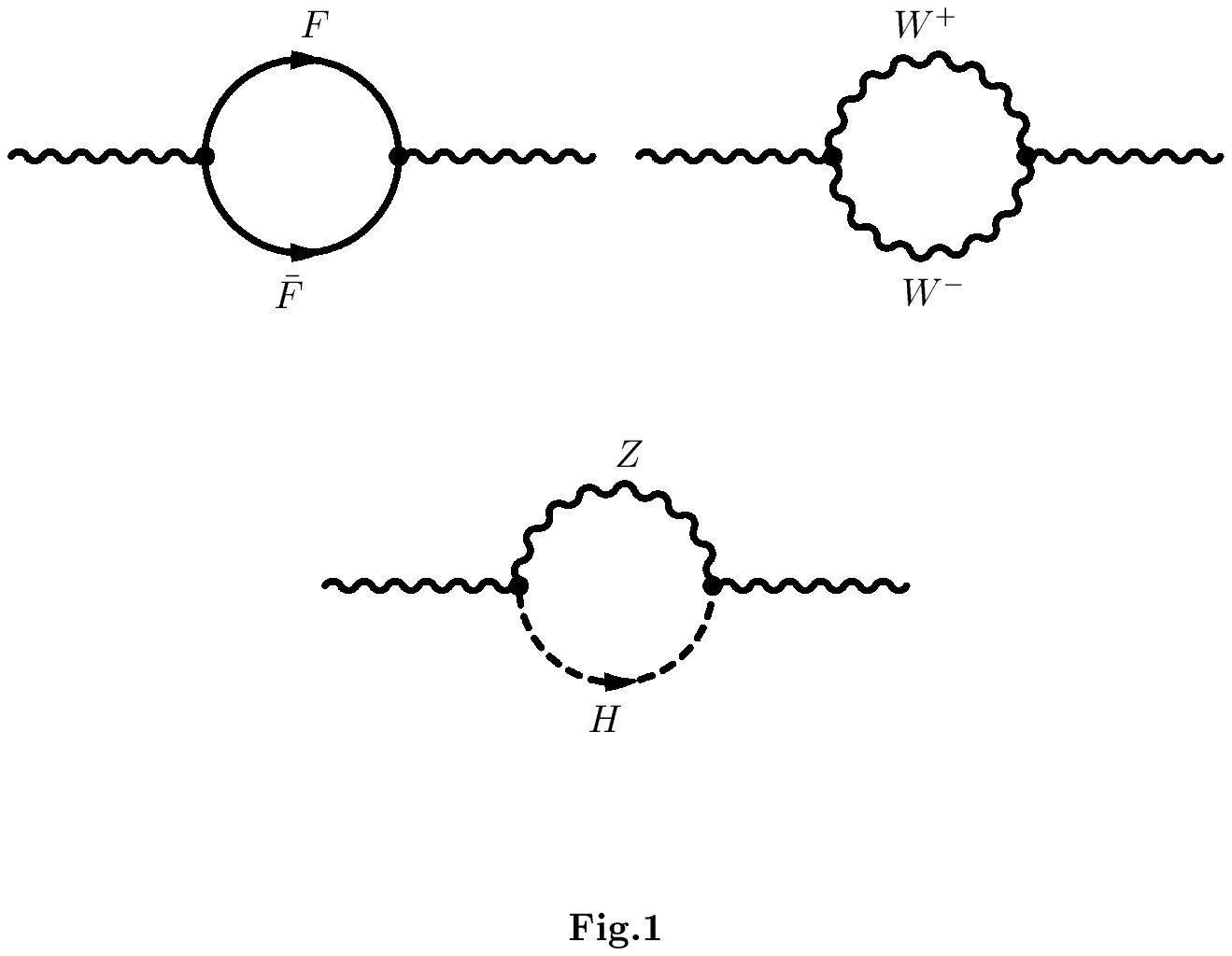,height=30cm}
\]
\vspace*{-15cm}
\caption[1]{Self-energy diagrams for $\gamma,Z$ gauge bosons. It must
be understood that $W$ or $Z$ running inside the loop are accompagned
by their corresponding Goldstones and ghosts states.}
\label{diag1}
\end{figure}
\begin{figure}[p]
\[
\hspace{-3cm}
\epsfig{file=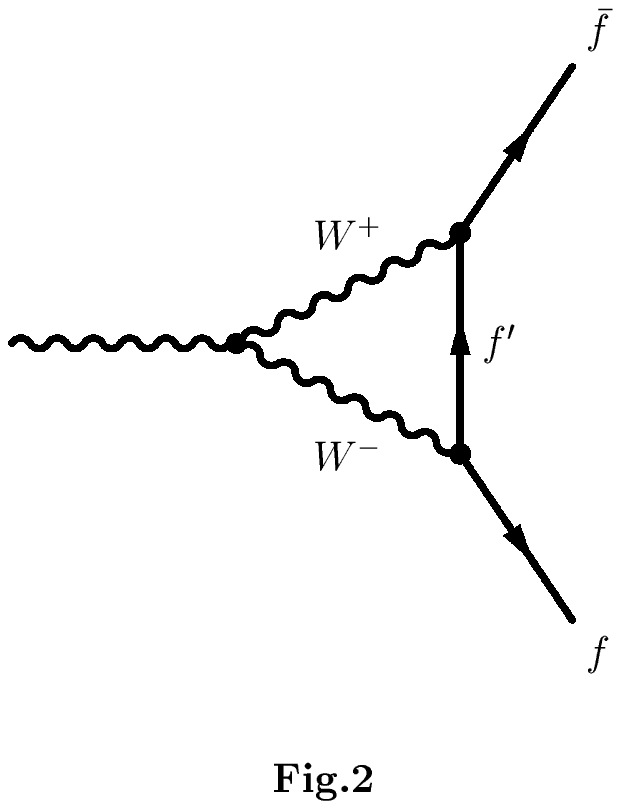,height=30cm}
\]
\vspace*{-18cm}
\caption[2]{$WW$ triangle contribution to the $\gamma-f\bar f$,
$Z-f\bar f$ vertex. Here also the contribution of the
corresponding Goldstones is to be added.}
\label{diag2}
\end{figure}
\begin{figure}[p]
\[
\hspace{-2.5cm}
\epsfig{file=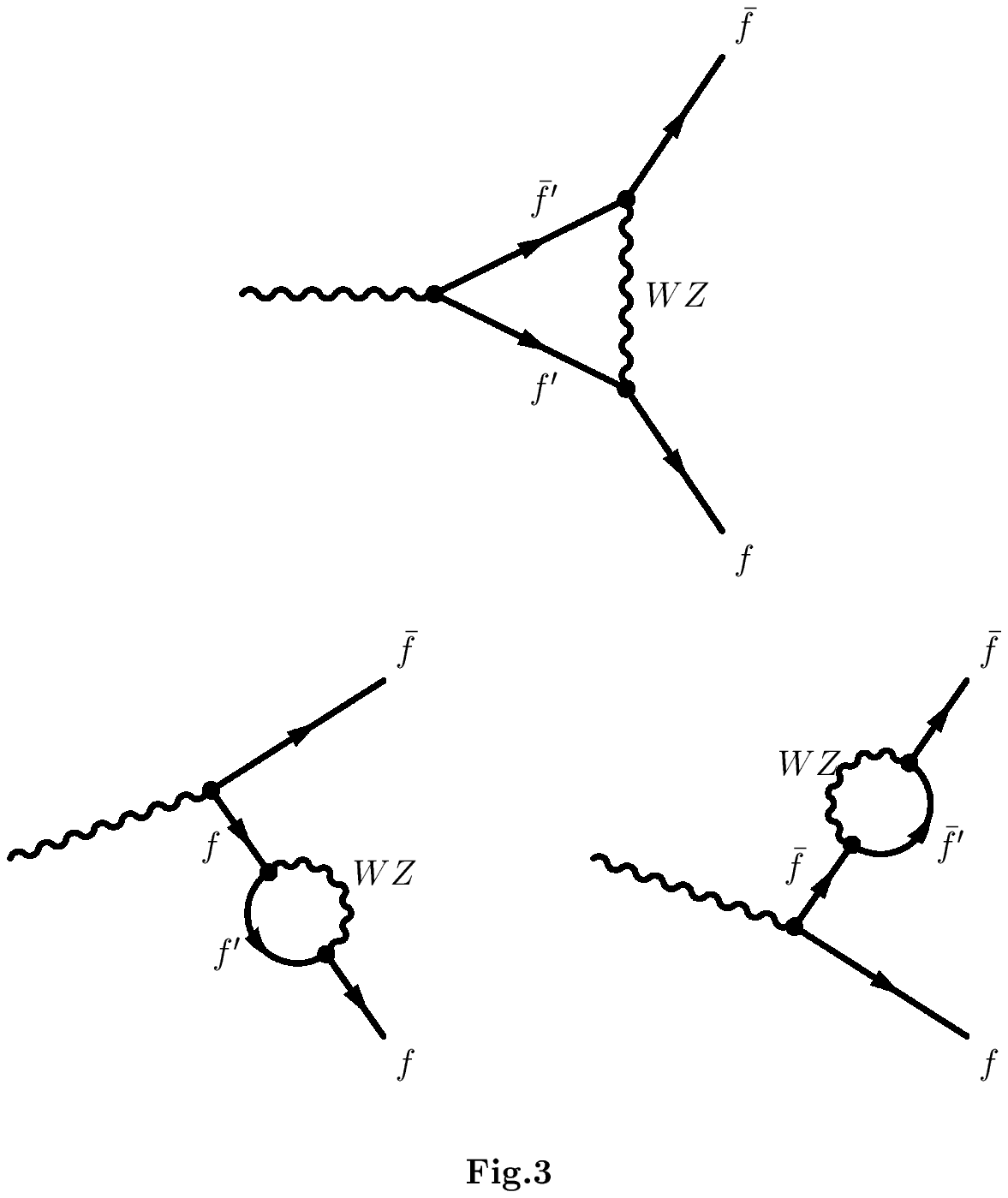,height=30cm}
\]
\vspace*{-13cm}
\caption[3]{Single $W$ or $Z$ exchange contribution
to the $\gamma-f\bar f$,
$Z-f\bar f$ vertex. Here also the contribution of the
corresponding Goldstone is to be added.}
\label{diag3}
\end{figure}
\begin{figure}[p]
\[
\hspace{-2.5cm}
\epsfig{file=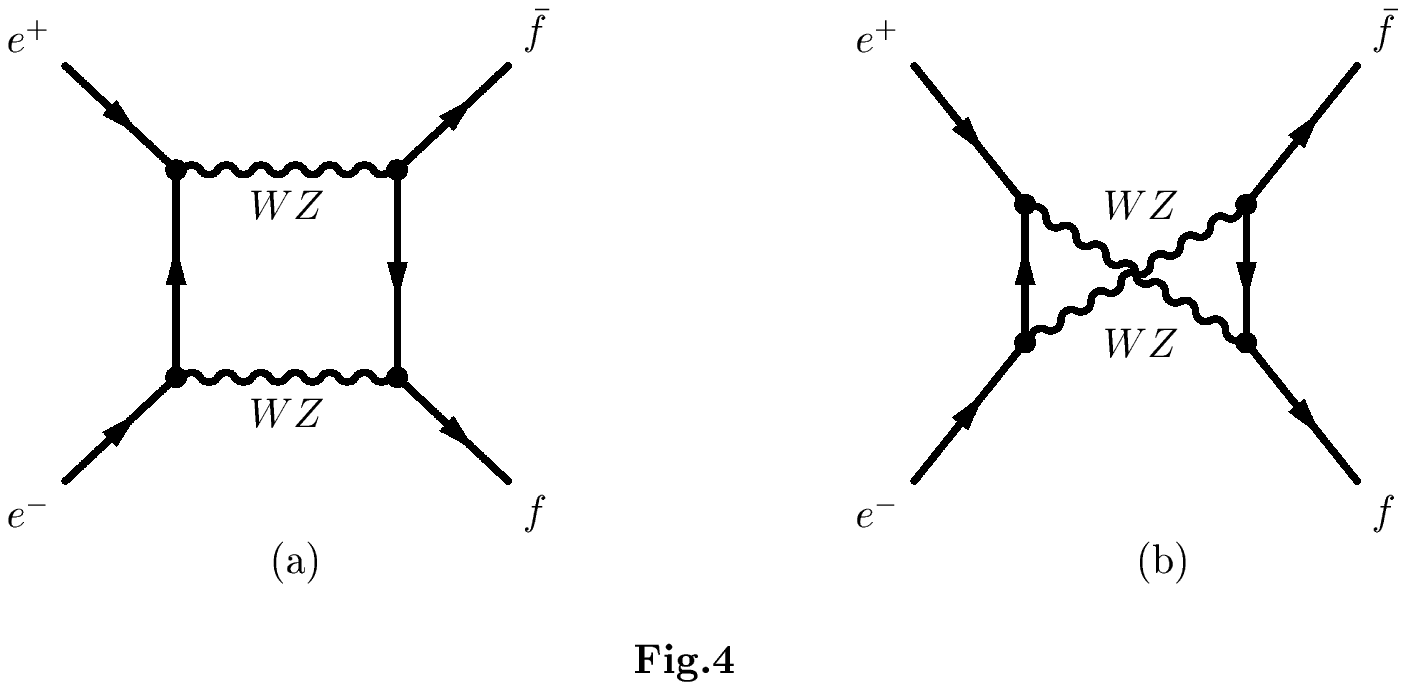,height=30cm}
\]
\vspace*{-19cm}
\caption[4]{$WW$ and $ZZ$ box contributions to $e^+e^-\to f\bar f$.
In the $WW$ case diagram (a) contributes for $I_{3f}=-{1\over2}$,
whereas diagram (b) contributes for $I_{3f}=+{1\over2}$. In the
$ZZ$ case both diagrams contribute.}
\label{diag4}
\end{figure}

\begin{figure}[p]
\[
\hspace{-2.5cm}
\epsfig{file=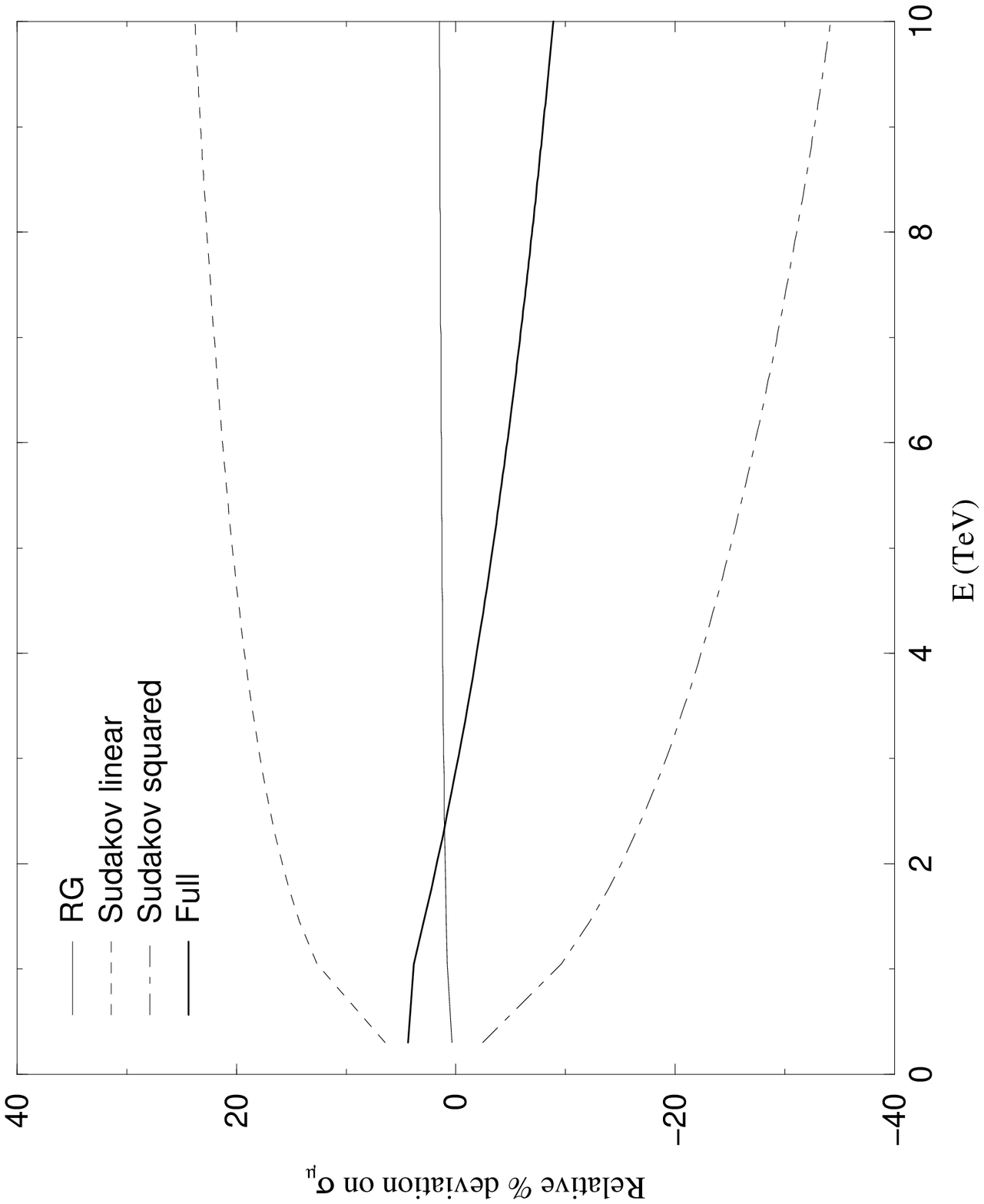,height=15cm}
\]
\caption[5]{Logarithmic contributions to the asymptotic
cross section $\sigma (e^+e^- \to \mu^+\mu^-)$ from
eq.(\ref{sigmu}).}
\label{Fig5}
\end{figure}
\newpage
\begin{figure}[p]
\[
\hspace{-2.5cm}
\epsfig{file=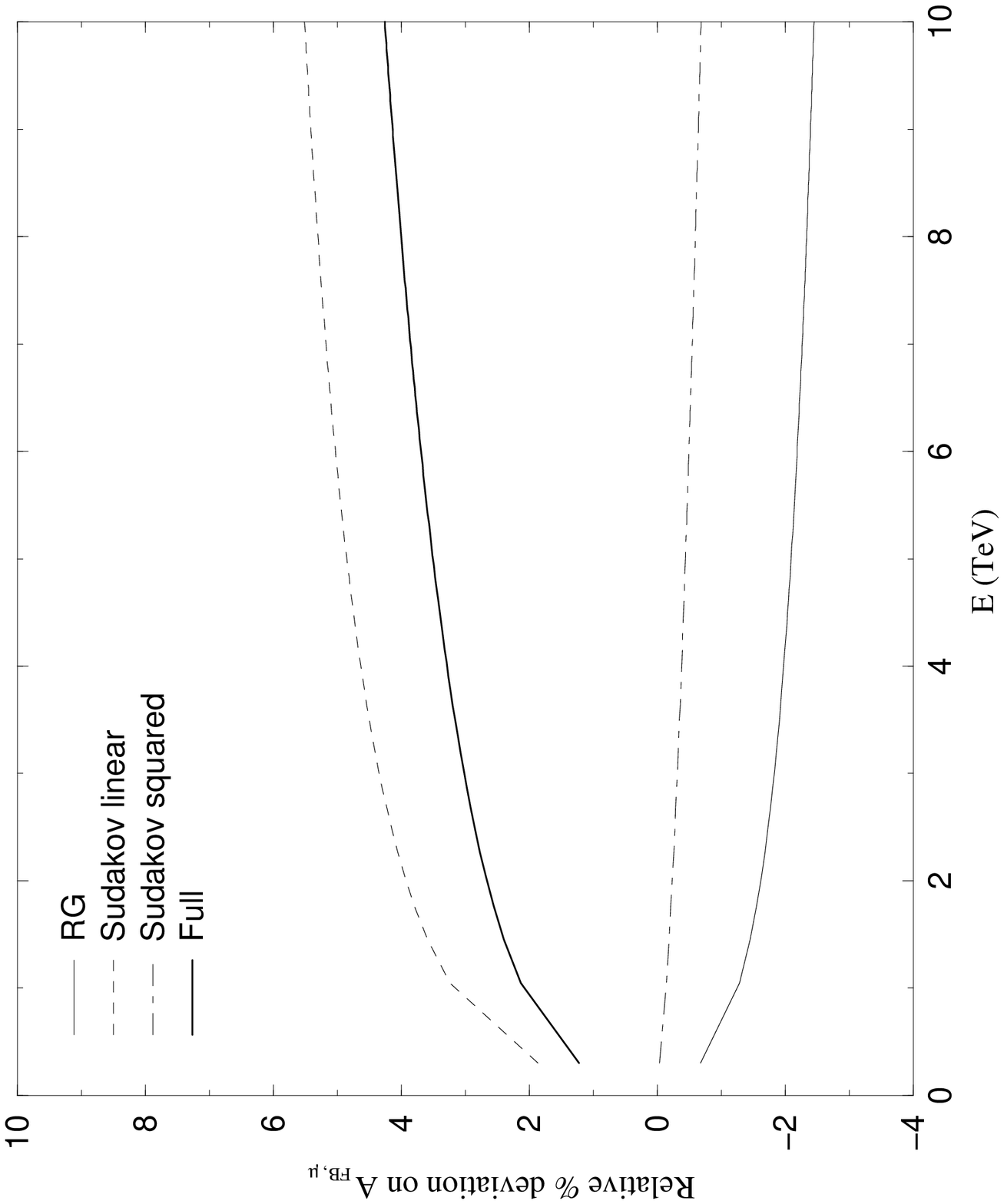,height=15cm}
\]
\caption[6]{Logarithmic contributions to the asymptotic
asymmetry $A_{FB,\mu}$ from
eq.(\ref{AFBmu}).}
\label{Fig6}
\end{figure}
\newpage
\begin{figure}[p]
\[
\hspace{-2.5cm}
\epsfig{file=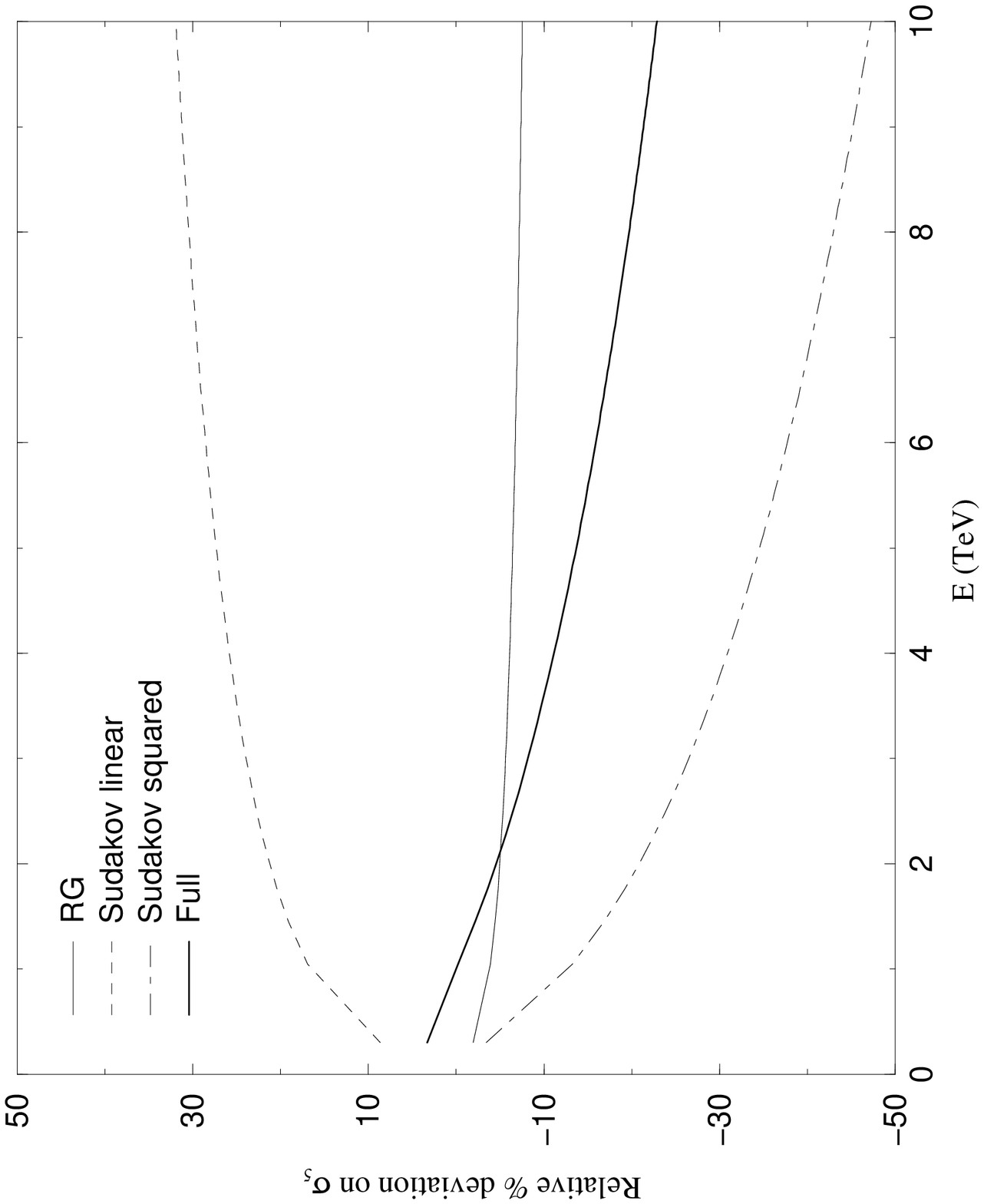,height=15cm}
\]
\caption[7]{Logarithmic contributions to the asymptotic
cross section $\sigma (e^+e^- \to hadrons)$ from
eq.(\ref{sigm5}).}
\label{Fig7}
\end{figure}
\newpage
\begin{figure}[p]
\[
\hspace{-2.5cm}
\epsfig{file=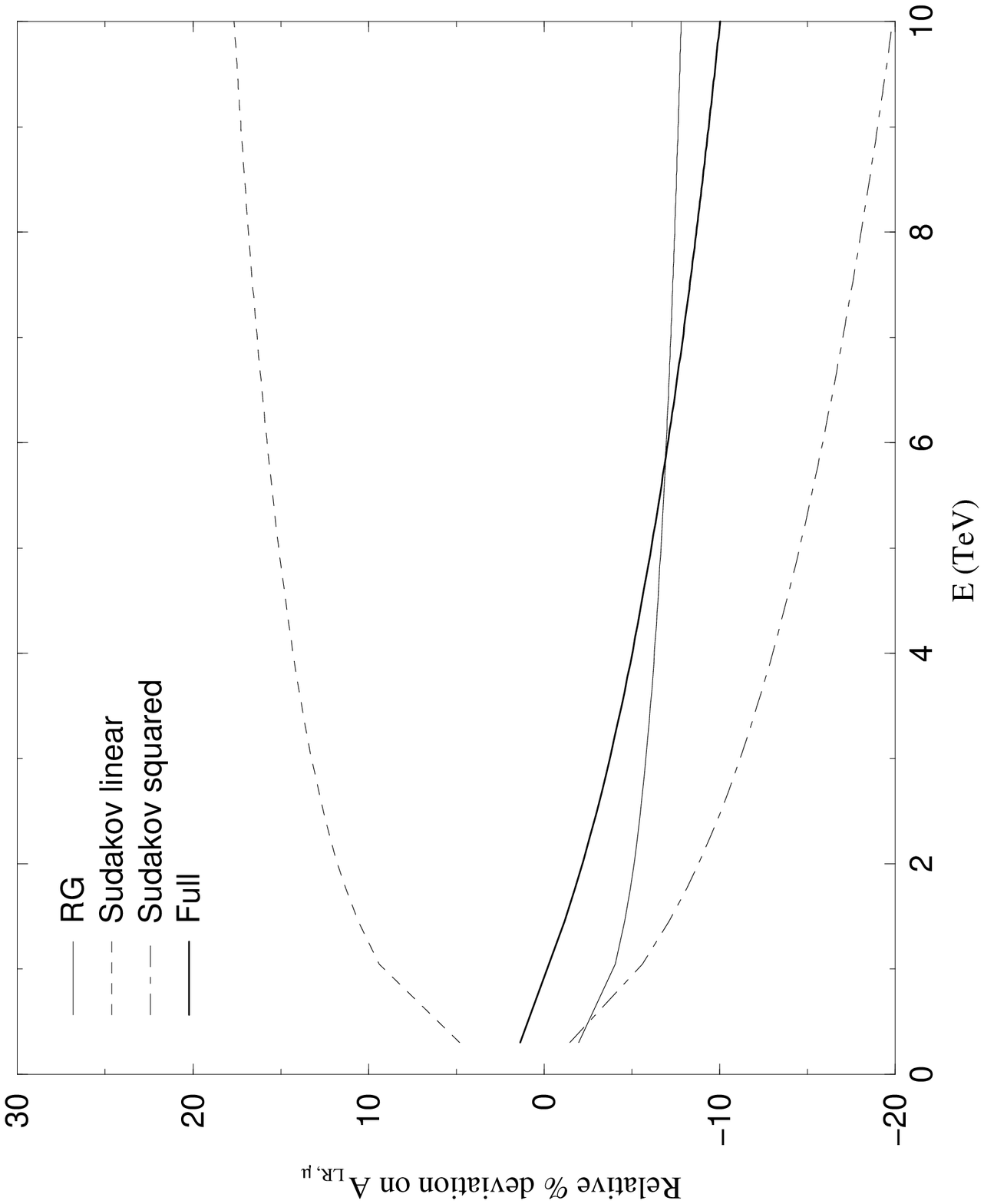,height=15cm}
\]
\caption[8]{Logarithmic contributions to the asymptotic
asymmetry $A_{LR,\mu}$ from
eq.(\ref{ALRmu}).}
\label{Fig8}
\end{figure}

\newpage
\begin{figure}[p]
\[
\hspace{-2.5cm}
\epsfig{file=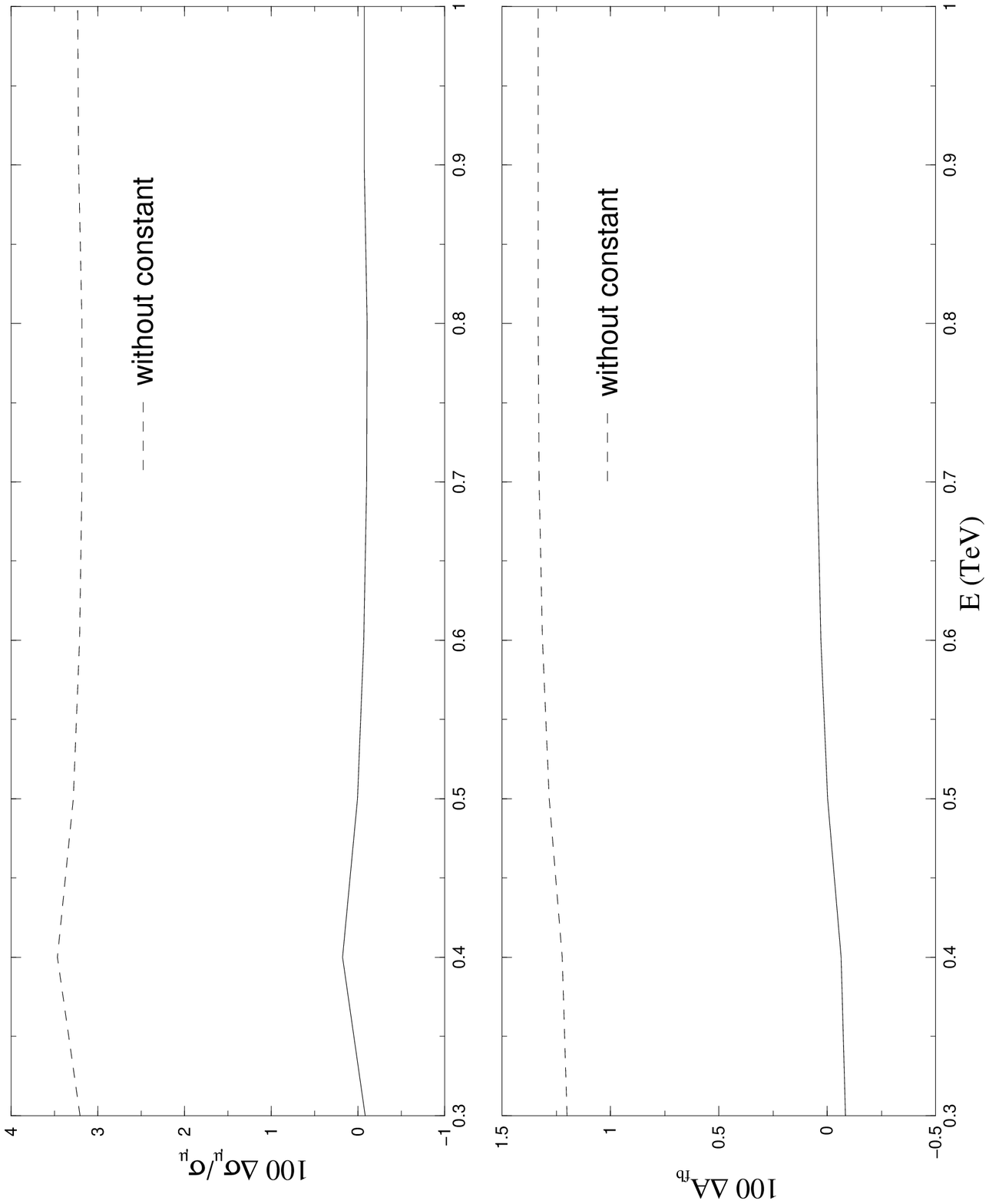,height=15cm}
\]
\caption[8]{Comparison between the asymptotic expressions for $\sigma_\mu$ and $A_{FB, \mu}$, Eqs.(\ref{sigm},
\ref{ass}) and the 
exact one loop calculation by TOPAZ0.}
\label{Fig9}
\end{figure}


\begin{thebibliography}{99}
\bibitem{data} A compact routine for the evaluation of the simple representations Eqs~(\ref{simple1},\ref{simple2}) 
including the complete Born expressions is available from the authors.
\end{thebibliography}
\end{document}